\documentclass[conference]{IEEEtran}
\IEEEoverridecommandlockouts
\usepackage{cite}
\usepackage{amsmath,amssymb,amsfonts}
\usepackage{algorithmic}
\usepackage{graphicx} 
\usepackage{textcomp}
\usepackage{xcolor}
\def\BibTeX{{\rm B\kern-.05em{\sc i\kern-.025em b}\kern-.08em
    T\kern-.1667em\lower.7ex\hbox{E}\kern-.125emX}}
\begin{document}

\title{A Stealthy Hardware Trojan Exploiting the Architectural Vulnerability of Deep Learning Architectures: Input Interception Attack (IIA)  \\
}

\author{\IEEEauthorblockN{1\textsuperscript{st} Tolulope A. Odetola}
\IEEEauthorblockA{\textit{Dept. of Electrical \& Computer Engr.} \\
\textit{Tennessee Technological University}\\
Cookeville, TN, USA 38505 \\
taodetola42@students.tntech.edu}
\and
\IEEEauthorblockN{2\textsuperscript{nd} Hawzhin Raoof Mohammed}
\IEEEauthorblockA{\textit{Dept. of Electrical \& Computer Engr.} \\
\textit{Tennessee Technological University}\\
Cookeville, TN, USA 38505 \\
hmohammed42@students.tntech.edu}
\and
\IEEEauthorblockN{3\textsuperscript{nd} Syed Rafay Hasan}
\IEEEauthorblockA{\textit{Dept. of Electrical \& Computer Engr.} \\
\textit{Tennessee Technological University}\\
Cookeville, TN, USA 38505 \\
shasan@tntech.edu}
}

\maketitle

\begin{abstract}
Deep learning architectures (DLA) have shown impressive performance in computer vision, natural language processing and so on. Many DLA make use of cloud computing to achieve classification due to the high computation and memory requirements. Privacy and latency concerns resulting from cloud computing has inspired the deployment of DLA on embedded hardware accelerators. To achieve short time-to-market and have access to global experts, state-of-the-art techniques of DLA deployment on hardware accelerators are outsourced to untrusted $3^{rd}$ parties. This outsourcing raises security concerns as hardware Trojans can be inserted into the hardware design of the mapped DLA of the hardware accelerator. We argue that existing hardware Trojan attacks highlighted in literature have no qualitative means how definite they are of the triggering of the Trojan. Also, most inserted Trojans show a obvious spike in the number of hardware resources utilized on the accelerator at the time of triggering the Trojan or when the payload is active.

In this paper, we propose a hardware Trojan attack called Input Interception Attack (IIA). In this attack we make use of the statistical properties of layer-by-layer output to make sure that asides from being stealthy, our IIA is able to trigger with some measure of definiteness.  This IIA attack is tested on DLA used to classify MNIST and Cifar-10 data sets. The attacked design utilizes approximately up to 2\% more LUTs respectively compared to the un-compromised designs. This paper also discusses potential defensive mechanisms that could be used to combat such hardware Trojans based attack in hardware accelerators for DLA.
\end{abstract}


\section{Introduction}
Deep learning architectures (DLA) is fast becoming a default standard in computer vision applications \cite{ma2018alamo}. DLA are designed with computationally intensive convolution operations \cite{hailesellasie2019mulnet} along with very memory centric multi-level perceptron layers.  \cite{fu2019machine}.

\begin{figure}[h]
\setlength{\abovecaptionskip}{0mm}   % 0.5cm as an example
\setlength{\belowcaptionskip}{0mm}   % 0`.5cm as an example
\centering
\includegraphics[width=0.5\textwidth]{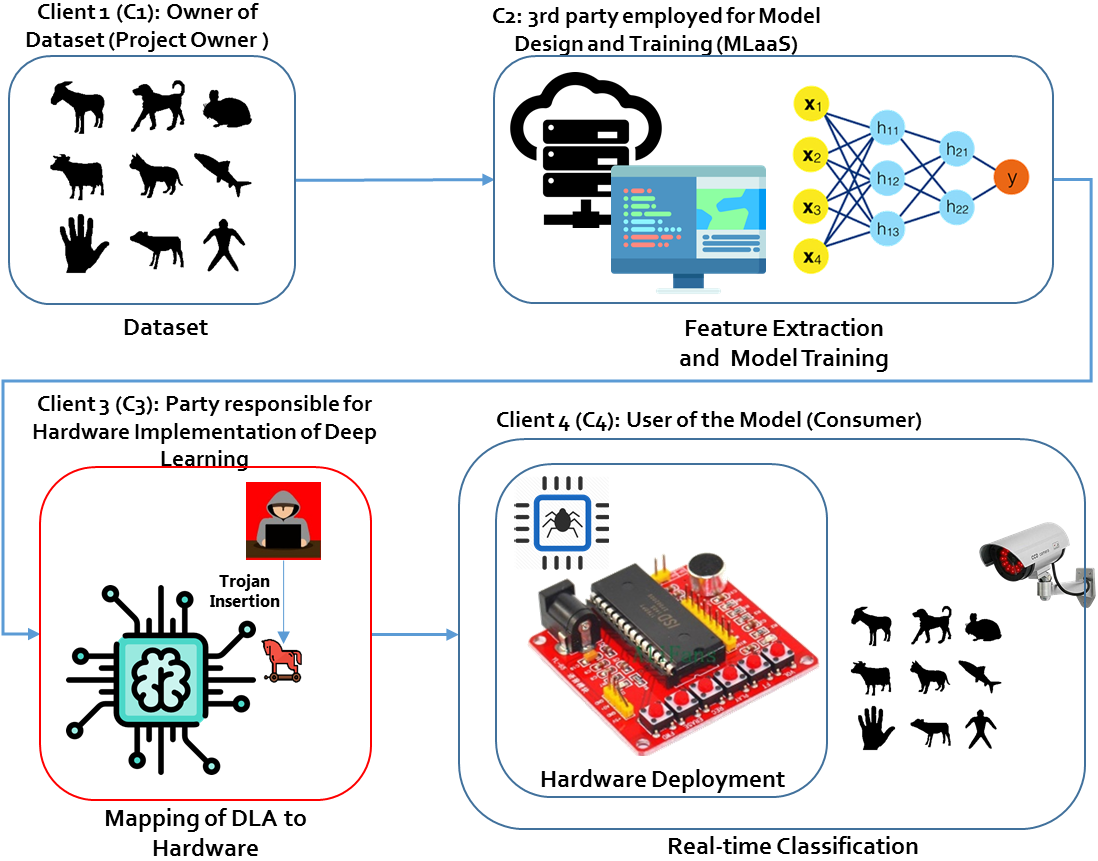}
\caption{Deep Learning Model Supply Chain}
\vspace{0mm}
\label{chain}
\end{figure}

DLA require huge quantity of training data to achieve good accuracy. The training of DLA is often outsourced to untrused third parties known as Machine Learning as a Service (MLaaS) as shown in Fig. \ref{chain}. This trained model then can be deployed into the hardware at another untrusted facility. Finally, this product is provided to the customer who can make use of such DLA based intelligent devices. 

%The high computational complexity of DNNs poses a formidable challenge in the deployment of DNNs on embedded, battery powered mobile devices, where resources are limited %or constrained. Utilizing FPGAs for the deployment of DNN models has attracted significant attention in recent times. FPGAs offers at low-precision computation, fast %prototyping and re-configuration which support rapidly changing DNN models \cite{yang2019synetgy}. Significant efforts has been made to achieve the deployment of DNN on %constrained devices. There is a strong potential that DNNs will be embedded in many consumer electronics, industrial and military equipment in the nearest future\cite{ %clements2018hardware}. This advancement gives rise to new security concerns.

Software or hardware Trojans can be inserted into the DLA by adversaries in any untrusted third party facilities of the supply chain. For instance, in the training phase, security concerns arise as $3^{rd}$ party model designers or trainers have access to training data sets, testing data sets, model architecture and model parameters (weights and biases). In situations where the model designer or trainer have malicious intent, malicious synapses or neurons not included in the original model design specifications can be hidden or embedded in the trained model to cause mis-classification of the model at a later time when triggered. \cite{zou2018potrojan}.

\begin{figure*}[t]
\centering
\setlength{\abovecaptionskip}{0mm}   % 0.5cm as an example
\setlength{\belowcaptionskip}{0mm}   % 0`.5cm as an example
\centering
\includegraphics[width=1.05\textwidth]{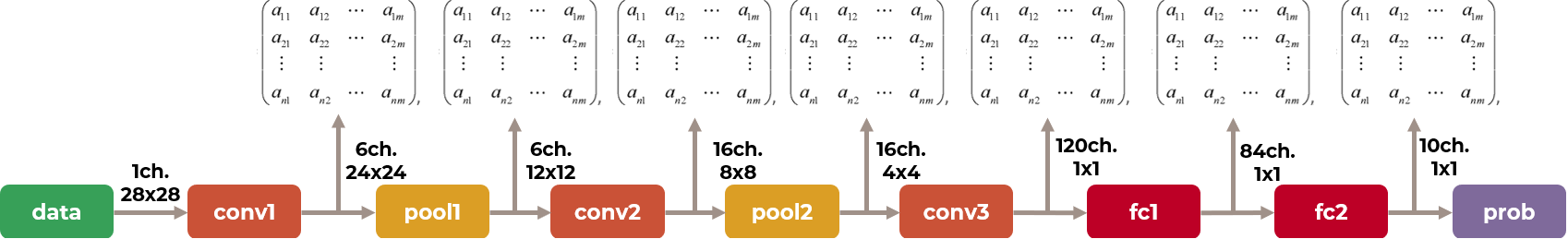}
\caption{DLA Gray Box Attack}
\vspace{0mm}
\label{concept}
\end{figure*}

  After training, many   DLA   models   make   use of  cloud  computing  to  achieve  classification during inference phase to overcome the  high computation  cost and  memory  requirements. Security, privacy  and  latency concerns  because of  cloud  computing  have  inspired  the  deployment  of  DLA  models  on  embedded  hardware  accelerators. To achieve short time-to-market and have access to global experts, current state-of-the-art  techniques  of  DLA  deployment on hardware accelerators, is also outsourced to untrusted $3^{rd}$ parties. Hardware Trojans can be inserted into hardware  accelerators for DLA architectures in such untrusted 3rd parties. Such Trojans can cause mis-classification  when  they  are  triggered as shown in Fig. \ref{chain}. 

Several hardware Trojan insertion techniques have been explored for hardware accelerators  for DLA architectures. Liu et. al in \cite{liu2017trojaning}  shows the possibility of a Trojan attack on neural networks that helps to generate samples of the input data from the pre-trained model. In this approach, they generate input trigger that can induce malicious behavior of the neurons in the model. Clements et. al \cite{clements2018hardware} introduces a hardware Trojan framework carried on IP designs. In this framework, input image trigger is used to generate small bounded perturbations that are added to targeted layers of the DLA to induce and propagate malicious behaviour through the DLA layers leading to mis-classification. 

These above approaches require a introduction of malicious functions or changing of the DLA parameters within the DLA layers which will incur more hardware resources when the Trojan is triggered and can consequently lead to detection. Also, the attacker has no qualitative guarantee that the Trojan will be triggered to achieve his objective because the trigger is designed based on input image trigger that may never be used by the end-user.  

In this work, we consider a DLA as shown in Fig. \ref{concept}. The hardware designer has access to the model architecture, the dimensions of the input matrix to the DLA and the dimensions of the respective layer-by-layer feature maps. With access to all these, the adversary can perform statistical analysis to obtain boundary conditions that qualitatively guarantees the Trojan triggering and also assures a high level of stealthiness.

In this paper, we propose a hardware Trojan attack called the Input Interception Attack (IIA). This attack does not require a change in parameters or introduction of malicious function within the DLA layers to cause mis-classification. This attack is designed to be perpetrated during the deployment of DLA on the hardware as shown in Fig. \ref{chain}. This Trojan attack is comprises of 3 phases namely: statistical analysis of the output features of each layer in DLA, the triggering circuit and payload. The IIA remains dormant unless it is triggered based on the statistical properties of certain output features of a particular layer (any layer can be chosen by the attacker) of DLA if these statistical properties satisfy the requirements for the trigger of the the IIA. Once the payload is activated the output of the DLA is compromised and consequently may lead to mis-classification. 

Our work  has following major contributions. :
\begin{itemize}
\item Propose a  novel and stealthy hardware Trojan attack whose triggers is dependent upon the layer-by-layer outputs of the DLA architecture 

\item We propose solution schemes that can help the prevention of insertion of hardware Trojans in the hardware implementation phase of DLA. 
\end{itemize}

The remainder of this paper is organized as follows: Section II describes the threat model. Section III discusses the proposed attack design. Section IV  shows experimental results and discussion. Section V discusses coutermeasures against hardware Trojans. Section VI provides related works and comparison with state-of-the-art and Section VII concludes the paper.

%%%%%%%%%%%%%%%%%%%%%%%%%%%%%%%%%%%%%%%%%%%%%%%%
\begin{figure*}[t]
\centering
\setlength{\abovecaptionskip}{0mm}   % 0.5cm as an example
\setlength{\belowcaptionskip}{0mm}   % 0`.5cm as an example
\centering
\includegraphics[width=1.05\textwidth]{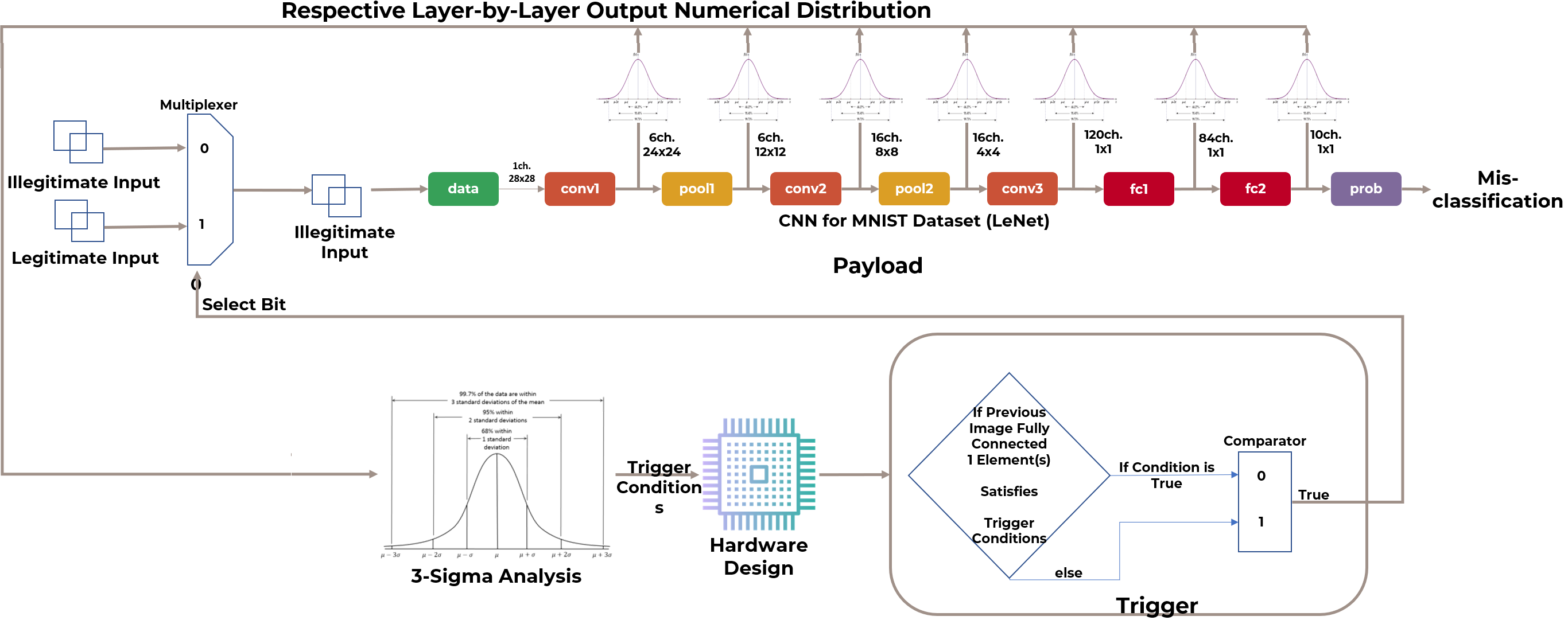}
\caption{Attack Design}
\vspace{0mm}
\label{trojan}
\end{figure*}

%%%%%%%%%%%%%%%%%%%%%%%%%%%%%%%%%%%%%%%%%%%%%%%%
\section{Threat Model}
The model adopted in this paper follows a Gray box attack where the adversary have some knowledge but not the full knowledge of the DLA architecture. The adversary is placed at the hardware implementation phase of deep learning architecture (DLA) as shown in Fig. \ref{chain}. In this model, the attacker has access to the DLA architecture, DLA parameters and a validation data set only from the project owner.  However, the attacker does not have any information about training and testing data samples. The validation data set is required to verify and validate the implementation correctness of the hardware design of the DLA. 

\section{Attack Design}
As stated earlier, the adversary has access to the parameters and architecture of the DLA. This means that the adversary has knowledge of the size of the input to the DLA and the layer-by-layer outputs of the DLA to any random input that follows the required size of the input to the DLA. The adversary also has access to a validation data set provided by project owner to help verify the correctness of the hardware design and implementation.

The IIA design is divided into three phases namely:
\begin{itemize}
\item Statistical Analysis
\item Trigger Design
\item Payload Design
\end{itemize}

%\begin{figure}[t]
%\setlength{\abovecaptionskip}{0mm}   % 0.5cm as an example
%\setlength{\belowcaptionskip}{0mm}   % 0`.5cm as an example
%\centering
%\includegraphics[scale=0.5]{DLAs.png}
%\caption{DLA Architecture for MNIST and Cifar-10 data sets}
%\vspace{0mm}
%\label{DLAs}
%\end{figure}

%%%%%%%%%%%%%%%%%%%%%%%%%%%%%%%%%%%%%%%%%%%%%%%%
\begin{figure}[t]
\centering
\setlength{\abovecaptionskip}{0mm}   % 0.5cm as an example
\setlength{\belowcaptionskip}{0mm}   % 0`.5cm as an example
\centering
\includegraphics[scale=0.45]{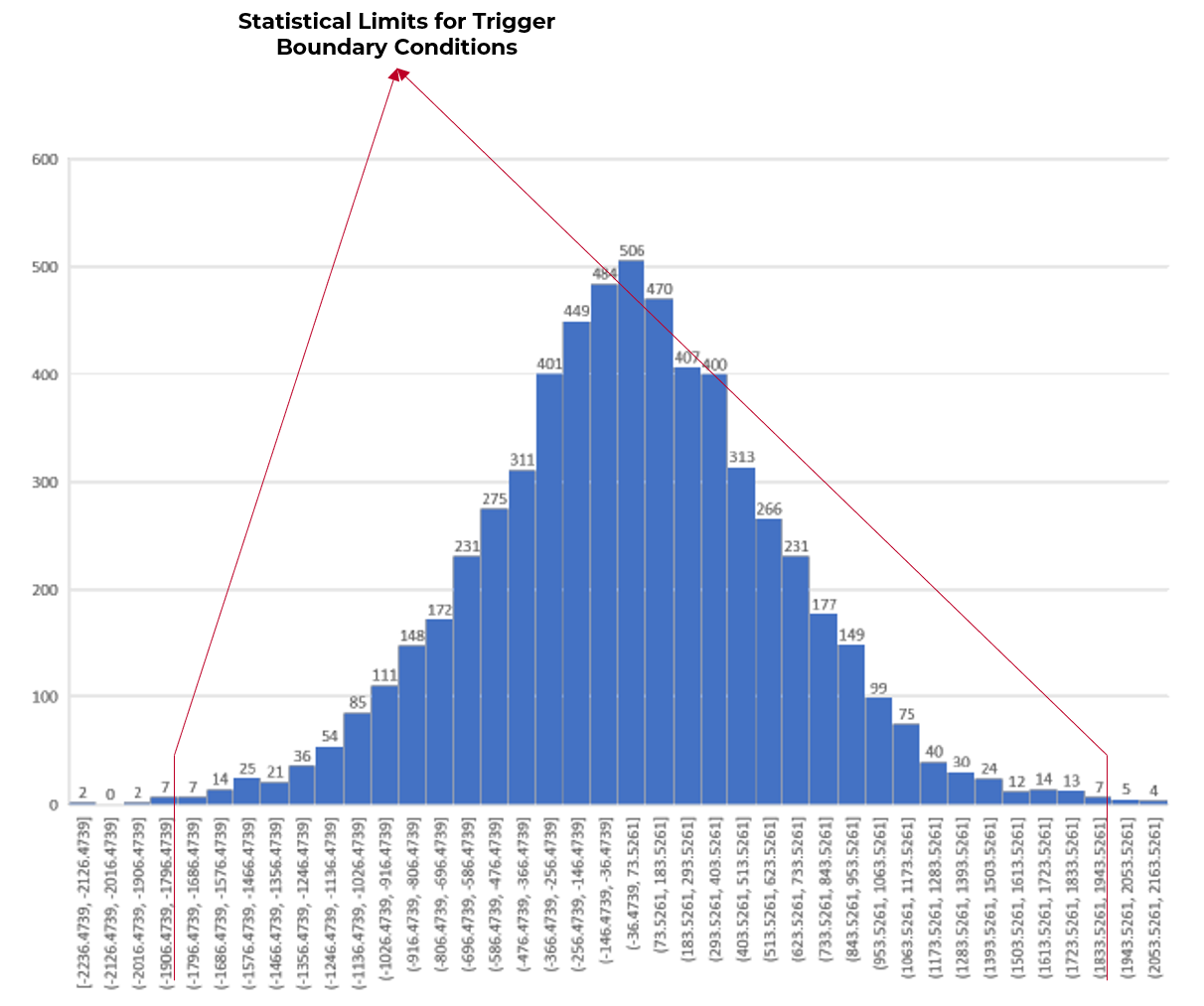}
\caption{Numerical Distribution of Fully Connected Layer 1 Elements}
\vspace{0mm}
\label{fig:fc1}
\end{figure}

%%%%%%%%%%%%%%%%%%%%%%%%%%%%%%%%%%%%%%%%%%%%%%%%
\subsection{Statistical Analysis}
After the hardware design, input images from the validation sets are used to test and verify the correctness of the hardware design simulations. The adversary has access to the respective layer-by-layer outputs of different input images to the DLA hardware design. In this stage, 3-sigma limit calculation is performed on the layer-by-layer outputs on validation data set to obtain the numerical distribution, frequencies and boundaries of the elements in the respective layer-by-layer outputs. To make the hardware Trojan stealthy but yet able to trigger this calculation helps in determining very rare but still possible range of output values that can be used for  effective trigger conditions.

From Fig. \ref{fig:fc1}, statistical limits of numbers ranging from 1833 to 1943 and also numbers ranging from (-1906) to (-1746) appears to be very rare and are suitable for stealthy and effective boundary conditions for the design of the trigger circuit.

\subsection{Trigger Design}
In this paper, statistical analysis is performed to obtain 3-sigma limit on the layer-by-layer output of the first fully connected layer shown in the DLA Fig. \ref{trojan}. This helped us obtain a range of values that are rarely possible to occur.  In the event when any element in the layer in question falls within those rarely occurring values, the Trojan is triggered. This paper assumes a validation data set of 100 images. These 100 images are used to simulate the hardware design to obtain numerical distribution of the first fully connected (fc1) layer of the DLA shown in Fig. \ref{trojan} over the 100 images of the validation data set. The numerical distribution of the fc1 layer is shown in Fig \ref{fig:fc1}. Numerical boundaries that lie between the 3-sigma and 4-sigma statistical limits are selected to design the trigger for the IIA. We called this range of values as the triggering boundary conditions in this paper.

\subsection{Payload Design}
In the event of an image satisfies the triggerring boundary conditions, the payload is designed to intercept and pass an illegitimate image as input in the next image cycle with the aim of causing mis-classification after which it resets itself and allows the passage of legitimate inputs to the design. This adversarial design stores the illegal images in a separate memory location. Since the number of these illegal images are very small therefore possible anticipated area overhead is very small. The saved image(s) can be made up of random or carefully selected pixel numbers that is passed through the DLA architecture and results in classification that deviates from the normal result leading to mis-classification. The payload takes action in the next image cycle after the trigger is activated. In order to make the payload more stealthy, we make sure that after mis-classification is achieved, the payload resets itself till the trigger is activated again. The Trojan design is summarized in Fig \ref{trojan}. The effectiveness of the attack is evaluated by measuring the following:
\begin{itemize}
\item Stealthiness: To measure the stealthiness of the attack, a large number of images from the given data set is passed through the compromised hardware design to observe the amount of times the proposed triggering mechanism is activated. 
\item Hardware resource utilization: This metric provides an understanding of how much additional hardware resources incurred as a result of the Trojan insertion in the IP design. This is measured by the number of Block-RAMs (BRAMs), Digital Signal Processors (DSPs), Flip-Flops (FFs), Look-Up Tables (LUTs) and hardware latency in comparison with a non-compromised hardware design of the DLA.
\end{itemize}

%%%%%%%%%%%%%%%%%%%%%%%%%%%%%%%%%%%%%%%%%%%%%%%%
\begin{table}[h]
\caption{HDRU for LeNet DLA}
\centering
\begin{tabular}{|c|c|c|c|c|c|}
\hline
\hspace{-5mm}    & \hspace{-2mm} BRAM \hspace{-2mm} & \hspace{-2mm} DSPs \hspace{-2mm} & \hspace{-2mm} Flip-Flops & LUTs \hspace{-2mm} & \hspace{-2mm} Latency (ms)\hspace{-2mm} \\
\hline
Original DLA    & 58   & 39   & 37750      & 42360   & 6.24205        \\ \hline
Compromised DLA & 58   & 39   & 37861      & 42741   & 6.24380        \\ \hline
\% Difference   & 0\%  & 0\%  & 0.294\%    & 0.895\% & 0.0280\%      \\ \hline
\end{tabular}
\label{lenet}
\end{table}

%%%%%%%%%%%%%%%%%%%%%%%%%%%%%%%%%%%%%%%%%%%%%%%%
\begin{table}[h]
\caption{HDRU for DLA on Cifar-10 Architecture}
\centering
\begin{tabular}{|c|c|c|c|c|c|}
\hline
\hspace{-5mm}    & \hspace{-2mm} BRAM \hspace{-2mm} & \hspace{-2mm} DSPs \hspace{-2mm} & \hspace{-2mm} Flip-Flops & LUTs \hspace{-2mm} & \hspace{-2mm} Latency(ms) \hspace{-2mm} \\
\hline
Original DLA    & 121  & 37   & 12311      & 16777 &  13.40099          \\ \hline
Compromised DLA & 121  & 37   & 12417      & 17141 &  13.40204           \\ \hline
\% Difference   & 0\%  & 0\%  & 0.857\%    & 2.146\%  & 0.0078\%          \\ \hline
\end{tabular}
\label{cifar}
\end{table}

%%%%%%%%%%%%%%%%%%%%%%%%%%%%%%%%%%%%%%%%%%%%%%%%

\section{Results and Discussion}

The IIA is implemented on DLA for Lenet and Cifar-10 architectures.  respectively. To measure the stealthiness of the attack, a total of 1000 images are passed through the hardware design. The Trojan is triggered in 20 image instances based on the chosen trigger conditions. The IIA has a 2\% trigger rate.

The Hardware Design Resource Utilization (HDRU) of the compromised hardware design is obtained and compared with the original hardware design. Tables \ref{lenet} and \ref{cifar} shows the different hardware resources and latency required by the original and the Trojan inserted hardware designs of the DLA for the Lenet and Cifar 10 architectures respectively.

The Table \ref{lenet} shows that the compromised HDRU uses approximately 1\% more LUTs, and 0.3\% more flip-flops,  compared to the original (uncompromised) hardware design for the LeNet architecture The compromised hardware design uses the same amount of hardware resources in terms of BRAM and DSPs when compared to the original DLA hardware design. The latency of the compromised hardware design is comparable as it is nominally increased by  0.0280\%.

For Cifar 10 architecture, the Table \ref{cifar} shows the HDRU  shows that the compromised design consumes 2\% more LUTs and 1\% more flip-flops compared to the original DLA hardware design which can also be considered tolerable. Again the compromised hardware design uses the same amount of hardware resources in terms of BRAM and DSPs when compared to the original DLA hardware design. The latency of the compromised hardware design increased very nominally to about  is 0.078\% more compared to the original DLA hardware design.

\section{Countermeasures: Defense Mechanism}
In this paper, a number of countermeasures are proposed to either detect or prevent the insertion of proposed Trojans. These countermeasures include:

\subsection{Altered Validation data set}
Based on the threat model adopted in this paper, the adversary (hardware designer) requires a validation data set to ensure the correct implementation of the DLA deployment on hardware accelerators. The validation data set consists of images used to confirm the prediction of the hardware design from client with the prediction of the software  trained model provided by the project owner. To ensure the validation data set does not serve as an avenue to compromise the hardware design, each image pixel is changed  scaling them randomly to make the respective images in the validation data set ambiguous and far from the actual image. This altered validation data set is sent to the hardware designer to validate the hardware design of deployment of DLA to the hardware accelerator. Hence, the adversary cannot have access to the actual layer-by-layer output to design an effective trigger. In the event the adversary designs a trigger based on the altered validation data set, the trigger conditions does not result in effective and stealthy attack.

For illustration, consider 

Matrix A of size $mxm$ representing the image data

Matrix B of size $nxn$ representing the a weight matrix. 

Matrix C of size $pxp$ representing the a output of convolution.

Such that:

\begin{align}
A \circledast B &= C
\end{align}

The altered image data sets, D is given as:
\begin{align}
r \cdot A &= D
\end{align}

where $r$ is a random and 

D is a matrix of size $nxn$ with ambiguous input pixels. Hence,

\begin{align}
D \circledast B &= C^{'}
\end{align}

where $C^{'}$ is an ambiguous output of randomness the adversary has access to.  Since $C$ and $C^{'}$ do not match, hence this leads to erroneous results for adversary and may also lead to Trojan which cannot be triggered.

\subsection{Distributed DLA Layer Implementation}
\begin{figure}[h]
\setlength{\abovecaptionskip}{0mm}   % 0.5cm as an example
\setlength{\belowcaptionskip}{0mm}   % 0`.5cm as an example
\centering
\includegraphics[scale=0.38]{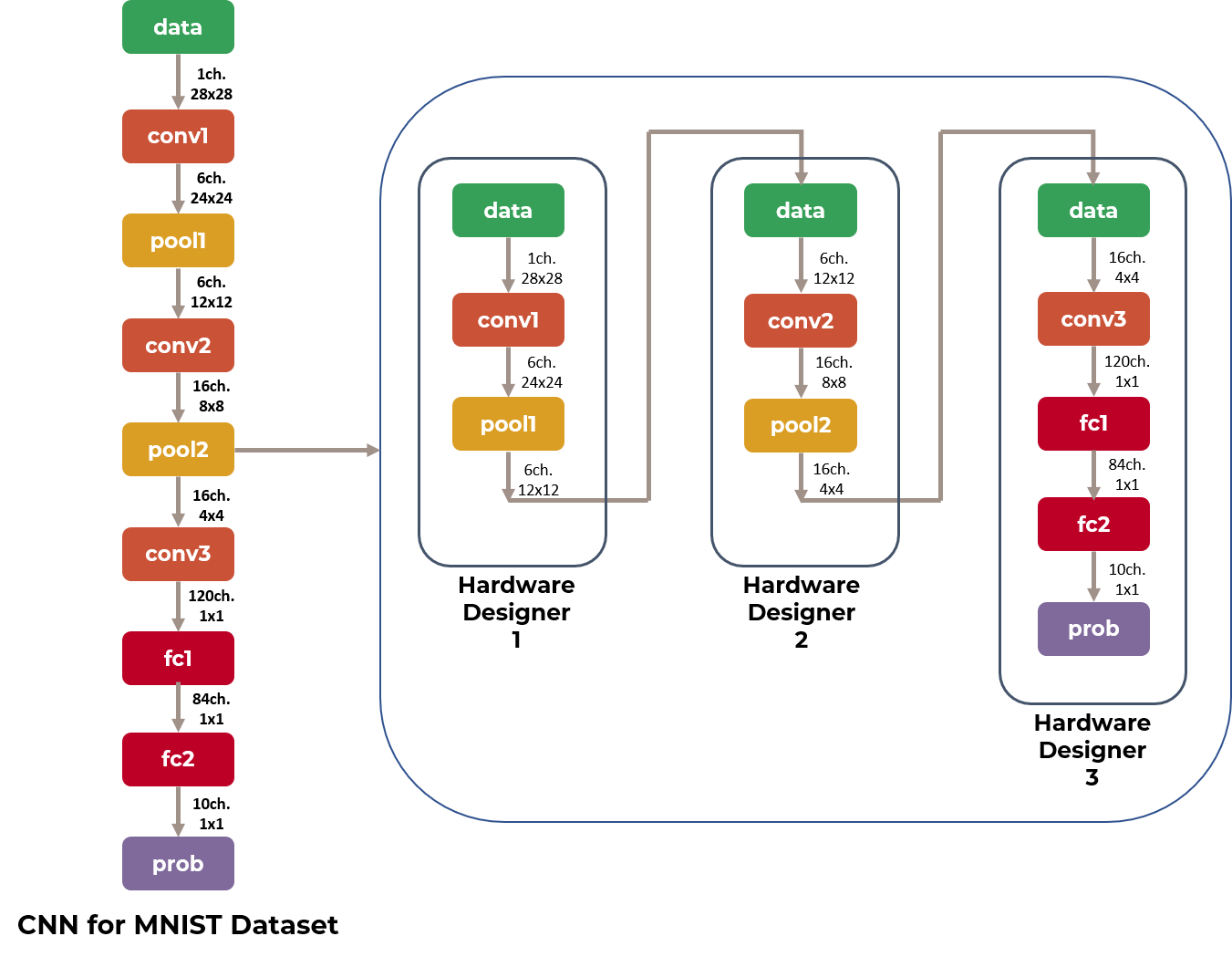}
\caption{Distributed DLA Layer Implementation}
\vspace{0mm}
\label{distributed}
\end{figure}

This defense mechanism proposes an approach where the software trained DLA model which is composed of layers are split into groups.
Each group contains a subset of the layers of the DLA model.
Each group is then contracted out to different non-related hardware designers to produce a standalone hardware design of the implementation of subset layers of the DLA.
Each hardware design are then put side by side to work together to perform classification.
This approach gives the adversary limited access to the parameters and the layer-by-layer output of the DLA model.
No adversary has complete knowledge of the actual size of the input and output to the overall DLA model.
This is illustrated in Fig \ref{distributed}.
The DLA model for LeNet architecture in Fig \ref{trojan} is split into groups and contracted to different hardware designers to limit the amount of knowledge that an individual hardware designer has access to in order to attack the model.

Hence, with the limited information, Trojan insertion is ineffective as the adversary has no means of measure the stealthiness or effectiveness of the attack.
This approach can also be used as a countermeasure for other form of attacks in literature as no adversary have access to full information about the DLA network.

%%%%%%%%%%%%%%%%%%%%%%%%%%%%%%%%%%%%%%%%%%%%%%%%

%%%%%%%%%%%%%%%%%%%%%%%%%%%%%%%%%%%%%%%%%%%%%%%%

%%%%%%%%%%%%%%%%%%%%%%%%%%%%%%%%%%%%%%%%%%%%%%%%
\section{Related Work}

Several schemes studied security and privacy~\cite{baza2019b,baza2018blockchain,baza2019blockchain,baza2019detecting,Park2019, pazos2019privacy}. Also, several applications have been developed to address security attacks in different applications~\cite{baza2,baza3,baza4,baza5,baza9,baza6,baza7,baza8,baza11}. Other approaches have been adopted to perform hardware Trojan attacks on pre-trained DLA implemented on hardware. Zou et al. \cite{zou2018potrojan} proposes a hardware Trojan on pre-trained models ready for hardware mapping called PoTrojan. In this approach, an input image is designed to be the trigger for the hardware Trojan. When the Trojan is triggered, the payload is designed to add minimal extra neurals and synapses to targeted layer(s) of the hardware design of the DLA. Without access to the training instances, the input image trigger is designed by using reverse engineering algorithms to obtain the layer-by-layer outputs of the DLA to obtain samples of the input image.
This approach requires additional hardware resource to accommodate additional synapses once the Trojan is triggered which shows a risk of its detection. The approach also requires Parameter Sensitivity Analysis (PSA) to determine the appropriate layer to add stealthy and effective malicious synapses that can be stealthy and effective.

Clements et. al \cite{clements2018hardware} proposes a hardware Trojan attack on neural networks that is targeted at specific layer(s). The attack is targeted at pre-trained neural networks that can be mapped to hardware accelerators. Here, adversarial images to fool the DLA are generated and used to determine the required perturbations needed to be added to the layer-by-layer output to cause mis-classification. In this approach, the Trojan is triggered with input images and upon triggering, the payload the malicious perturbations are added to the feature maps of the layers of the DLA. This approach requires additional hardware resource to accommodate addition of perturbations once the Trojan is triggered which shows a risk to its detection. The approach also requires PSA to determine the appropriate layer to add perturbations to that will make the attack stealthy and effective.

Clements et. al \cite{clements2019hardware} proposes a hardware Trojan design on neural networks that is targeted at computational blocks or activation function of specific layer(s) to cause mis-classification. This approach also tampers with the layers of the DLA to cause malicious behavior. Their approach requires layer-by-layer PSA to determine the appropriate layer and activation function that can be attacked to achieve stealthy and effective attack.

Zhao et. al \cite{zhao2019memory} proposes a memory Trojan attack on hardware accelerators. In this approach, the Trojan is located in the memory controller and is triggered by specific input images. After the Trojan is triggered, the payload inserts error in the layer-by-layer feature maps to cause mis-classification. This approach also tampers with the layer-by-layer feature map of the DLA to cause the malicious behavior. To achieve this layer-by-layer PSA to determine the appropriate layer that can be attacked to achieve stealthy and effective attack. The addition of erroneous data will require additional hardware resource which can lead to detection.

\begin{table}[h]
\centering
\caption{Summary of Comparison With State-of-the-Art Techniques}
\begin{tabular}{|c|c|c|c|c|c|}
\hline
Metrics                                            & \cite{zou2018potrojan} & \cite{clements2018hardware} & \cite{clements2019hardware} & \cite{zhao2019memory} & IIA                       \\ \hline
Designed without PSA      & x                                       & x                                            & x                                            & x                                      & \checkmark \\ \hline
Require no change in DLA parameters          & x                                       & x                                            & x                                            & x                                      & \checkmark \\ \hline
Require no change in activation functions    & x                                       & x                                            & x                                            & x                                      & \checkmark \\ \hline
Require no additional hardware after trigger & x                                       & x                                            & x                                            & x                                      & \checkmark \\ \hline
\end{tabular}
\label{compare}
\end{table}

These above state-of-the-art approaches in the related work are compared with the proposed methodology. The Table of comparison is shown in Table \ref{compare}.

%In the Parameter manipulation approach, using this approach, the adversary has limited %information and cannot decipher how relevant a parameter is before altering it

%%%%%%%%%%%%%%%%%%%%%%%%%%%%%%%%%%%%%%%%%%%%%%%%

\section{Conclusion and Future Works}
In this paper, we propose a hardware Trojan attack called the Input Interception Attack (IIA). This attack performs a statistical analysis on the layer-by-layer output of DLA to obtain stealthy trigger conditions for the design of the Trojan. This attack is a Gray box attack that is carried out in the hardware implementation stage of DLA architecture. This attack is demonstrated on two DLA used to perform classification on MNIST and Cifar-10 data sets. The IIA proved to be stealthy and adds negligible hardware overhead. This paper also proposes potential defense mechanisms to help combat this attack and other Trojan insertion techniques.

\bibliographystyle{IEEEtran}
\bibliography{TT.bib}
\end{document}